\input{aipcheck}

\documentclass[
    ,final            % use final for the camera ready runs
  ]
  {aipproc}

\layoutstyle{6x9}

\begin{document}

\title{X-ray observations of IC\,348 \newline in light of an updated cluster census}


\author{B. Stelzer}{
  address={INAF - Osservatorio Astronomico di Palermo, Piazza del Parlamento 1, I-90134 Palermo, Italy}
}
\author{G. Micela}{
  address={INAF - Osservatorio Astronomico di Palermo, Piazza del Parlamento 1, I-90134 Palermo, Italy}
}
\author{E. Flaccomio}{
  address={INAF - Osservatorio Astronomico di Palermo, Piazza del Parlamento 1, I-90134 Palermo, Italy}
}
\author{S. Sciortino}{
  address={INAF - Osservatorio Astronomico di Palermo, Piazza del Parlamento 1, I-90134 Palermo, Italy}
}

\begin{abstract}
IC\,348 is 
%a prototype of low-mass clustered star formation and 
an excellent laboratory for studies of low-mass star formation being
nearby, compact and rich. A {\em Chandra} observation was carried out early in the satellite's lifetime.
The extensive new data in optical and infrared wavelengths accumulated in subsequent years have changed the
cluster census calling for a re-analysis of the X-ray data. 
\end{abstract}

\maketitle

%%%%%%%%%%%%%%%%%%%%%%%%%%%%%%%%%%%%%%%%%%%%
%% MAINMATTER
%%%%%%%%%%%%%%%%%%%%%%%%%%%%%%%%%%%%%%%%%%%%

\section{The current IC\,348 cluster census}

IC\,348 is the nearest ($\sim 310$\,pc), rich and compact ($d \sim 20^{\prime}$)
cluster of low-mass star formation.
To date, more than $300$ 
members in a wide mass range ($\sim 0.02-5\,M_\odot$) 
have been spectroscopically confirmed, most of them within the central cluster area of $16^\prime \times 14^\prime$. 
The extinction is relatively low ($A_{\rm V} < 4$\,mag for most known members) and the mean cluster age 
is $2$\,Myr. 

With these characteristics IC\,348 is a prime laboratory for studies of low-mass star formation. 
A $53$\,ksec of IC\,348 was
presented by \cite{Preibisch01.1, Preibisch02.1}. Subsequently, extensive work was done on the cluster
at optical and infrared wavelengths: 
\begin{itemize}
\item $\sim 130$ new cluster members were discovered by signatures of youth in optical low-resolution spectra and their
position in the HR diagram above the $10$\,Myr isochrone \cite{Luhman03.2}; 
\item the stellar parameters have been revised shifting some objects to 
earlier spectral type, i.e. higher mass \cite{Luhman03.2}; 
\item Young Stellar Objects (YSOs) have been classified using {\em Spitzer} photometry \cite{Lada06.1}; 
\item rotation periods were measured for $> 100$ cluster members \cite[e.g.][]{Cieza06.1}.
\end{itemize}

\begin{figure}
    \parbox{0.5\textwidth}{
     \includegraphics[width=0.49\textwidth]{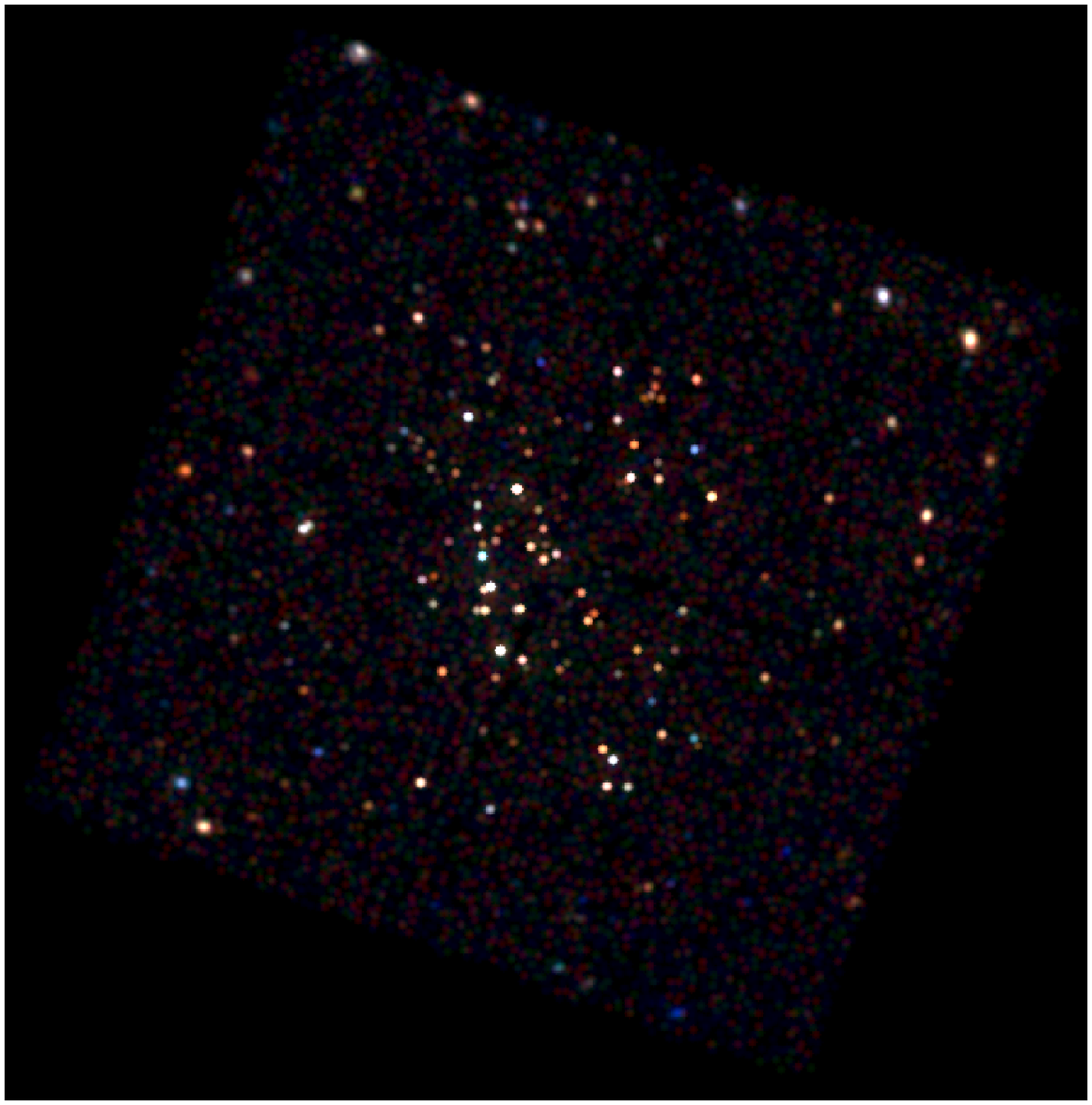}
    }
    \parbox{0.5\textwidth}{
     \includegraphics[width=0.49\textwidth]{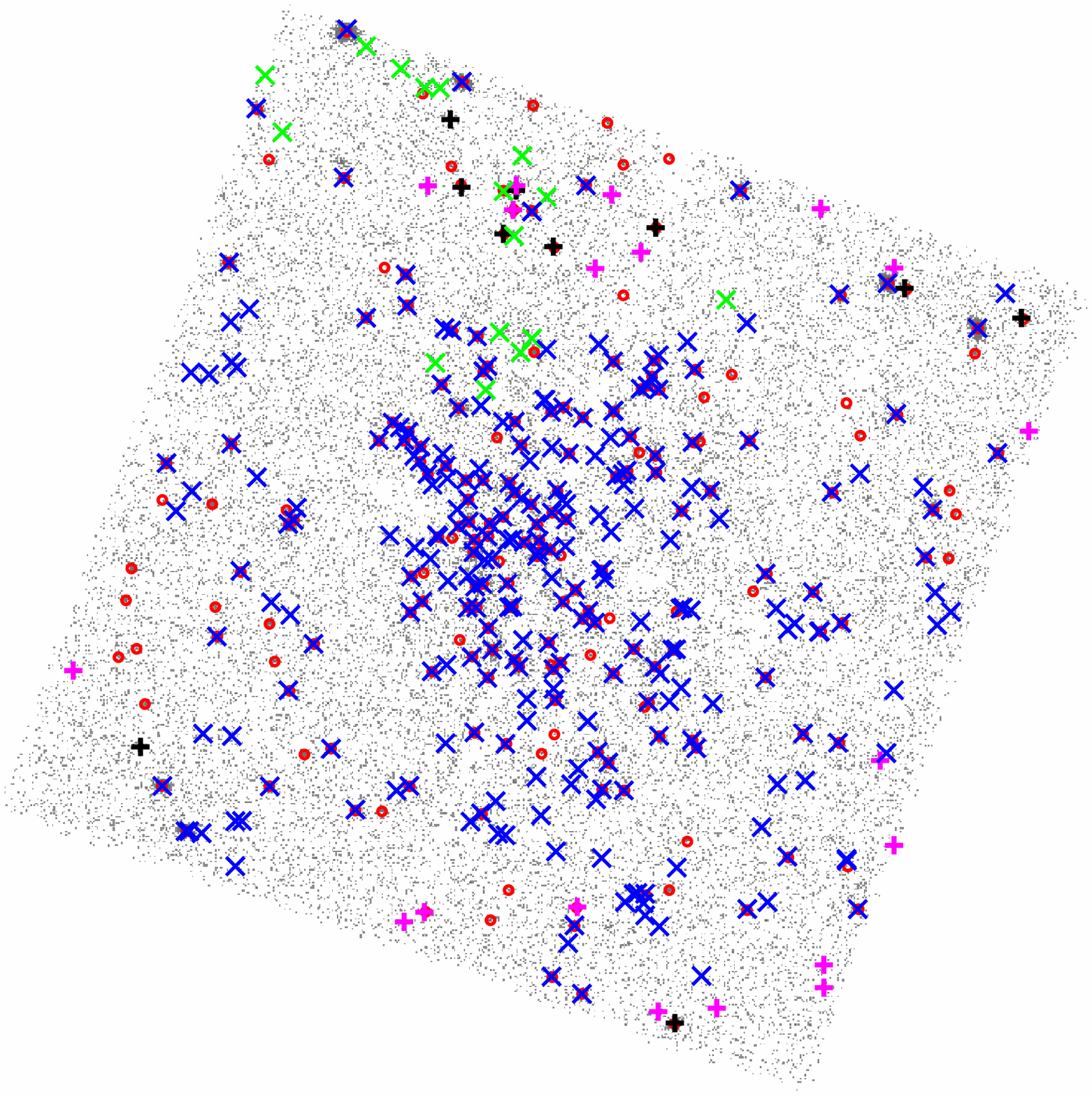}
    }
  \caption{{\em Chandra}/ACIS-I image of IC\,348:  
(LEFT) - false-color image; (RIGHT) - 
image with X-ray sources from \protect\cite{Preibisch01.1, Preibisch02.1} (red circles)
and cluster members from the literature: blue \protect\cite{Luhman03.2}, green \protect\cite{Luhman05.7},
pink and black \protect\cite{Muench07.1}; blue, green and pink sources are spectroscopically confirmed IC\,348 members.}
  \label{fig:1}
\end{figure}

All these stellar properties impact on the interpretation of the X-ray data, such as the relation between
X-ray luminosity and stellar parameters ($L_{\rm bol}$, $T_{\rm eff}$, $M$), X-ray luminosity functions,
the rotation-activity relation, etc. Therefore, we present a re-analysis of the {\em Chandra}
observation of IC\,348 incorporating the updated optical and infrared database. 
Two representations of the {\em Chandra} image are shown in Fig.~\ref{fig:1}.

\section{Data analysis}

In this paper we use the results from our new analysis of the X-ray data  
only for known IC\,348 members absent from the list
of detections in \cite{Preibisch01.1, Preibisch02.1}.
For all detections in common with our new analysis the X-ray data from \cite{Preibisch01.1, Preibisch02.1} is used. 
The X-ray source list was cross-correlated with the IC\,348 master catalog compiled from the references
given in Sect.~1. 
The re-analysis is also used to extract upper limits for undetected IC\,348 members. 
This is mandatory for a complete X-ray census, because upper limits are not available 
from \cite{Preibisch01.1, Preibisch02.1} for the IC\,348 members discovered after 2001.

\section{Results}

\subsection{X-rays and evolutionary stage}

IC\,348 is rich in Young Stellar Objects (YSOs) of all evolutionary stages. 
Class\,I protostars, disk-bearing 
Class\,II T Tauri stars and disk-less Class\,III T Tauri stars can be distinguished 
by the slope of their spectral energy distribution (SED) in the mid-IR. 
The spectral index $\alpha_{\rm SED}$ derived from 
{\em Spitzer} mid-IR photometry for IC\,348 members was presented by \cite{Lada06.1}. 
We assign the YSO class to each X-ray source using the boundaries for 
$\alpha_{\rm SED}$ defined by \cite{Lada06.1} and \cite{Muench07.1}. 

The X-ray detection statistics for the different YSO classes in IC\,348 
can be summarized as follows: $N_{\rm det}/N_{\rm ul} = 5/7$ (Class\,I),
$45/55$ (Class\,II), $36/31$ (Class\,II/III), and $93/22$ (Class\,III) where $N_{\rm det}$ is the number
of detections and $N_{\rm ul}$ the number of undetected stars.

\subsection{X-ray luminosity functions}

We computed X-ray luminosity functions (XLF) for different YSO classes (to examine the influence of disks) and for
different levels of H$\alpha$ emission (to examine the influence of accretion). To avoid biases related
to the known dependence of $L_{\rm x}$ on stellar mass, the analysis was carried out in three mass bins.
%The significance of the results is verified with $2$-sample tests that yield the probability for the
%observed luminosity distributions to be drawn from different parent distributions. 
%Both XLF and $2$-sample tests are computed within the ASURV\footnote{ASURV...} environment. 
The XLF take into account the upper limits for undetected stars. 

\begin{figure}
\includegraphics[width=0.33\textwidth]{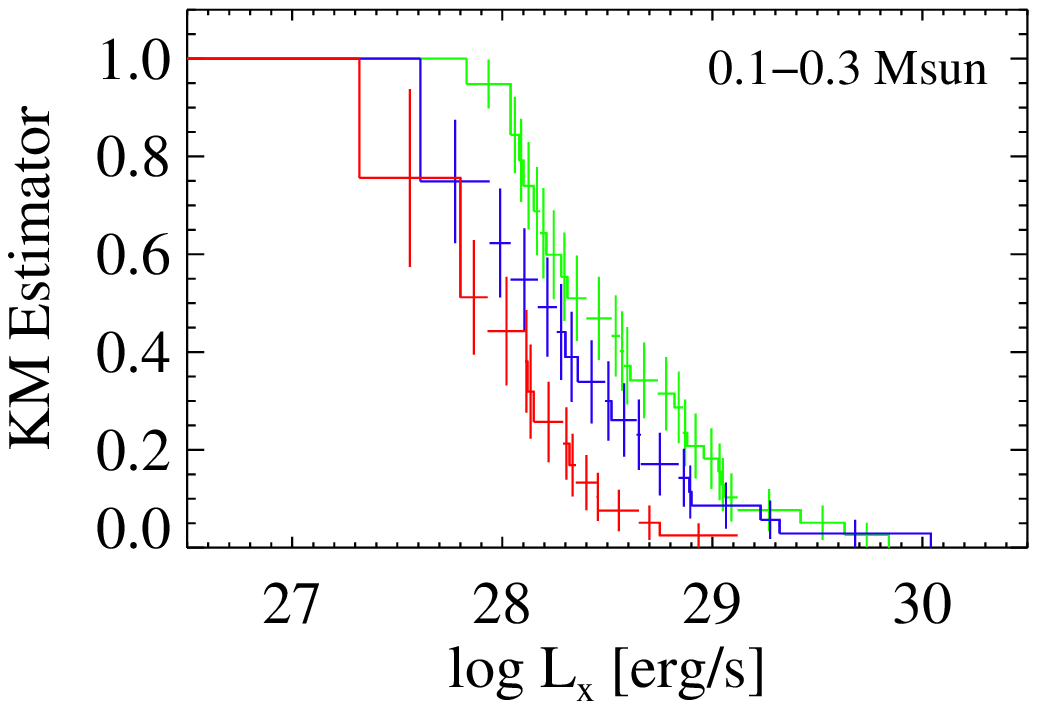}

\includegraphics[width=0.33\textwidth]{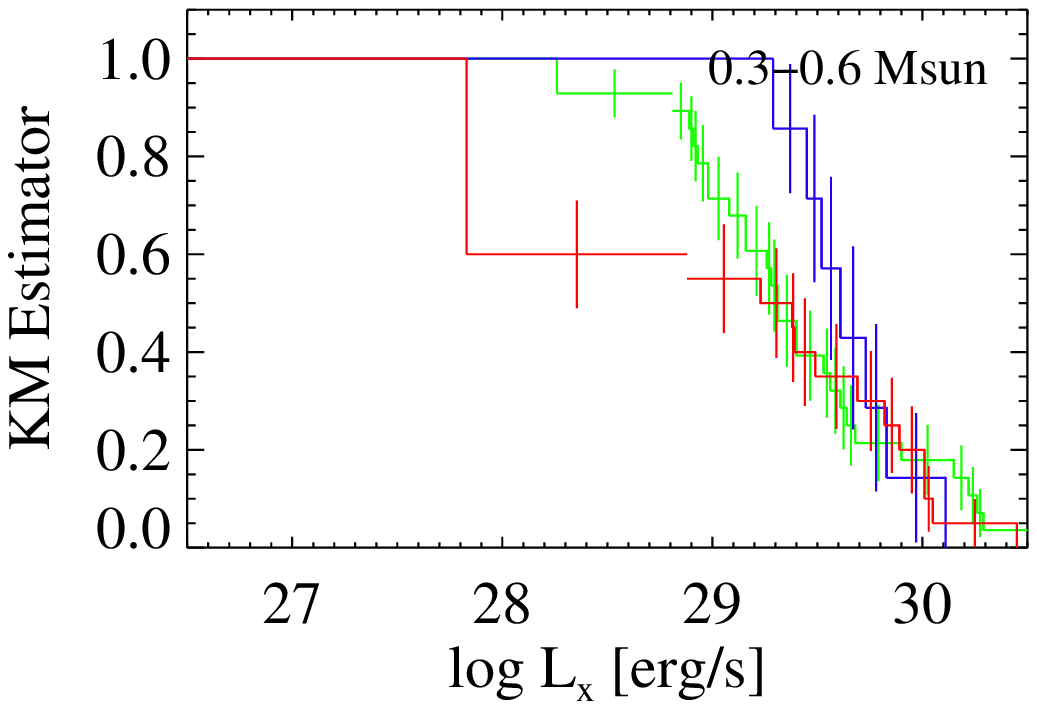}

\includegraphics[width=0.33\textwidth]{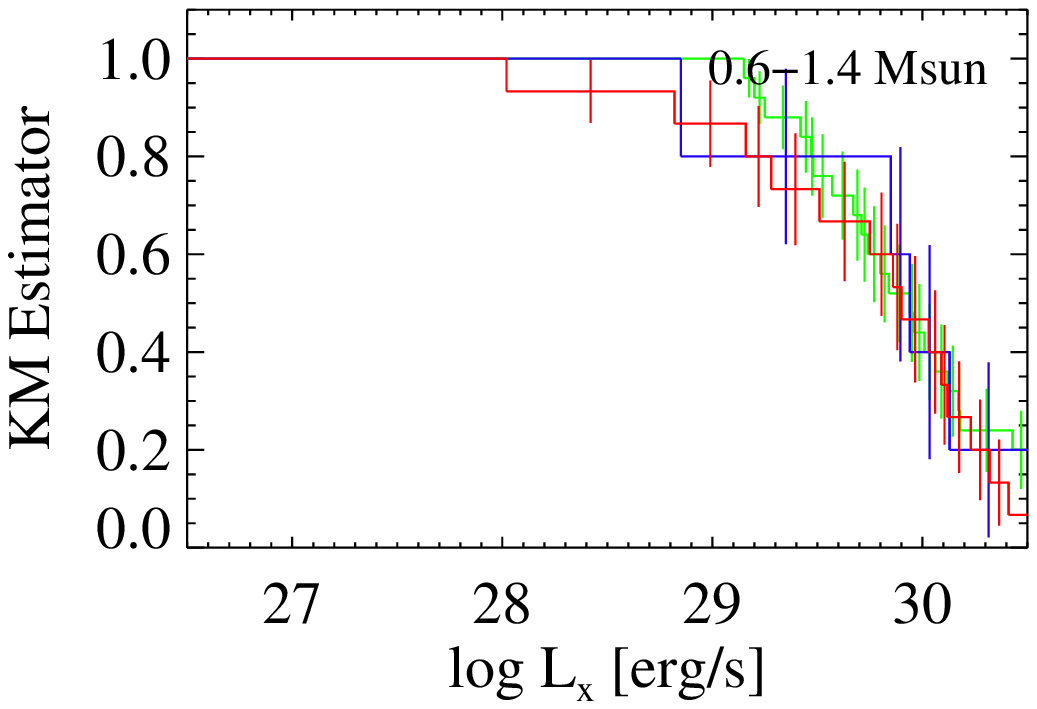}
\caption{X-ray luminosity functions for the {\em Spitzer} YSO classes in different mass bins. {\em red} - Class\,II,
{\em blue} - Class\,II/III, {\em green} - Class\,III.}
\label{fig:2}
\end{figure}

In Fig.~\ref{fig:2} we compare the XLF of Class\,II, II/III, and III sources. Class\,I objects are not considered because
of poor statistics. There is a pronounced difference in the X-ray luminosities of the three YSO classes for
the lowest mass bin ($0.1-0.3\,M_\odot$), while for higher mass stars the XLF of the three types of YSOs can not
be distinguished.
%The smaller X-ray luminosities of Class\,II with respect
%to more evolved Class\,II/III and Class\,III sources is statistically significant. 
%

\subsection{X-rays and stellar parameters}

\begin{figure*}[t]
\parbox{16cm}{
\parbox{8cm}{
\includegraphics[width=8cm]{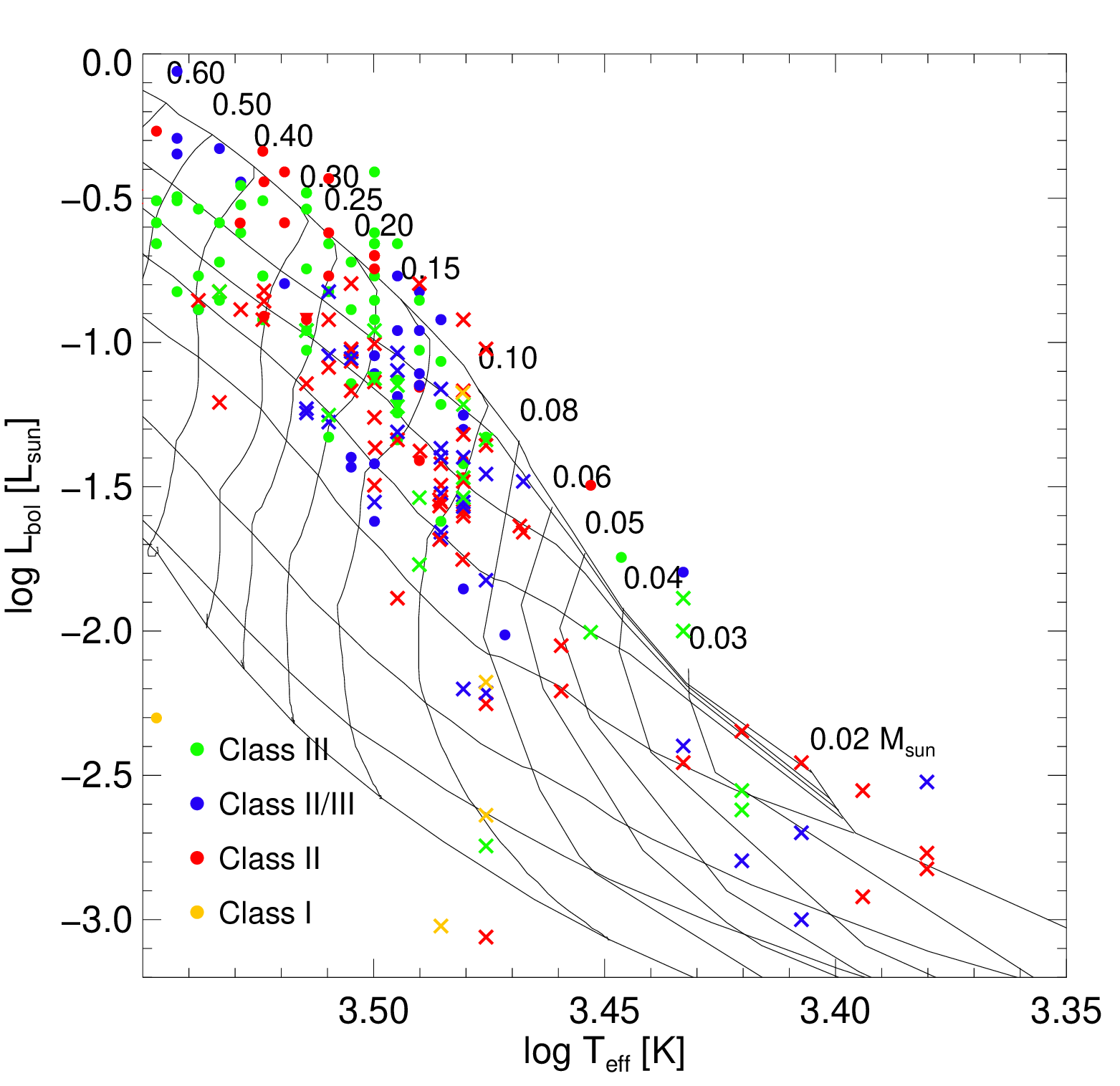}
}
\parbox{8cm}{
\includegraphics[width=8cm]{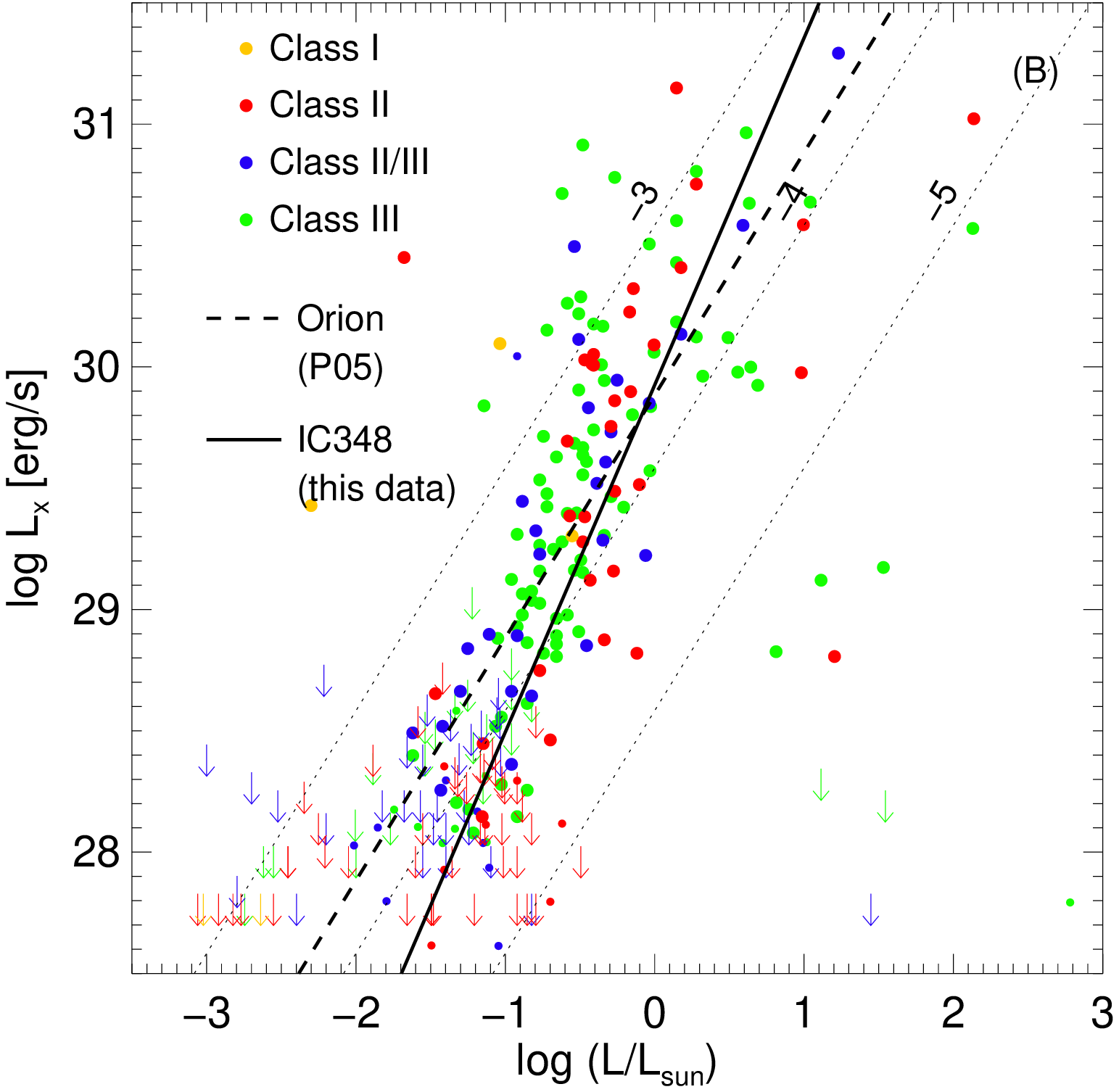}
}
}
\caption{
\newline
(LEFT) -- Low-mass region of the HR diagram for IC\,348 compared to evolutionary tracks of
\cite{Baraffe98.1} and \cite{Chabrier00.2}. 
{\em filled circles} - X-ray detections, {\em crosses} - non-detections; 
%the objects are color-coded according to their YSO class (see text). 
\newline
(RIGHT) -- X-ray versus bolometric luminosity; dotted lines represent constant $L_{\rm x}/L_{\rm bol}$
ratios $10^{-3}$, $10^{-4}$ and $10^{-5}$; the dashed line is the linear regression fit derived 
by \cite{Preibisch05.1} for the Orion Nebula Cluster 
corresponding to $\log{(L_{\rm x}/L_{\rm bol})} = -3.7$; the solid line is a linear regression fit
to the IC\,348 members with $M \leq 1.2 M_\odot$ excluding some flaring objects. 
} 
\label{fig:3}
\end{figure*}

Fig.~\ref{fig:3} shows the HR diagram and the $L_{\rm x}$ vs. $L_{\rm bol}$ diagram for IC\,348. 
Evidently most undetected cluster members are very-low mass stars and brown dwarfs. 
The sample of known IC\,348 members extends to lower
bolometric luminosities and masses than that of Orion, possibly due to both environmental and observational effects
(e.g. higher extinction in Orion).
Therefore, IC\,348 is excellent for
studying the relation between X-ray activity an stellar parameters in very-low mass stars and brown dwarfs.
Contrary to what was observed for Orion and Taurus \cite[see ][]{Preibisch05.1, Grosso07.1}, 
we find that the relation between $L_{\rm x}$ and $L_{\rm bol}$ in IC\,348 seems to 
be steeper than expected for constant $L_{\rm x}/L_{\rm bol}$ ratio (see Fig.~\ref{fig:3}~{\em right}).
This is due to a large number of X-ray faint (undetected) low-mass stars. Note, that the fitted curve for
IC\,348 includes only stars with $> 0.1\,M_\odot$ where upper limits are well constrained.
%It is not clear if this is
%an effect of a dependence of $L_{\rm x}/L_{\rm bol}$ on mass, on effective temperature, 
%or bolometric luminosity itself. 
%

Clearly, at the low-luminosity end the X-ray data are dominated by insufficiently constrained upper limits. 
Some hypotheses for a possible decrease of $L_{\rm x}/L_{\rm bol}$ level towards lower masses are: 
%This mass range ($0.3...0.1\,M_\odot$) is probably critical for the 
(i) transition from a solar-like to a convective dynamo, combined with a lower efficieny of the latter one 
resulting in decreased X-ray production;
(ii) smaller co-rotation radius for lower-mass stars and the ensuing 
centrifugal disruption of the corona \citep{Jardine06.1}; 
(iii) a mass dependent fraction of stars with X-ray emission suppressed by accretion 
\citep{Flaccomio03.2}. 
A deeper X-ray observation is necessary to unveil the behavior of the $L_{\rm x}/L_{\rm bol}$ level
for the faint and cool objects with the lowest stellar masses ($0.1...0.3\,M_\odot$).

%%%%%%%%%%%%%%%%%%%%%%%%%%%%%%%%%%%%%%%%%%%%%%%%
%% BACKMATTER
%%%%%%%%%%%%%%%%%%%%%%%%%%%%%%%%%%%%%%%%%%%%%%%%

%\begin{theacknowledgments}
%\end{theacknowledgments}



\bibliographystyle{aipproc}   % if natbib is available

\bibliography{ms}

\begin{thebibliography}{13}
\expandafter\ifx\csname natexlab\endcsname\relax\def\natexlab#1{#1}\fi
\providecommand{\enquote}[1]{``#1''}
\expandafter\ifx\csname url\endcsname\relax
  \def\url#1{\texttt{#1}}\fi
\expandafter\ifx\csname urlprefix\endcsname\relax\def\urlprefix{URL }\fi
\providecommand{\eprint}[2][]{\url{#2}}

\bibitem[Preibisch and Zinnecker(2001)]{Preibisch01.1}
T.~Preibisch, and H.~Zinnecker \textbf{122}, 866--875 (2001).

\bibitem[{Preibisch} and {Zinnecker}(2002)]{Preibisch02.1}
T.~{Preibisch}, and H.~{Zinnecker}, \emph{\aj} \textbf{123}, 1613--1628 (2002).

\bibitem[{Luhman} et~al.(2003)]{Luhman03.2}
K.~L. {Luhman}, J.~R. {Stauffer}, A.~A. {Muench}, G.~H. {Rieke}, E.~A. {Lada},
  J.~{Bouvier}, and C.~J. {Lada}, \emph{\apj} \textbf{593}, 1093--1115 (2003).

\bibitem[{Lada} et~al.(2006)]{Lada06.1}
C.~J. {Lada}, A.~A. {Muench}, K.~L. {Luhman}, L.~{Allen}, L.~{Hartmann},
  T.~{Megeath}, P.~{Myers}, G.~{Fazio}, K.~{Wood}, J.~{Muzerolle}, G.~{Rieke},
  N.~{Siegler}, and E.~{Young}, \emph{\aj} \textbf{131}, 1574--1607 (2006).

\bibitem[{Cieza} and {Baliber}(2006)]{Cieza06.1}
L.~{Cieza}, and N.~{Baliber}, \emph{\apj} \textbf{649}, 862--878 (2006).

\bibitem[{Luhman} et~al.(2005)]{Luhman05.7}
K.~L. {Luhman}, E.~A. {Lada}, A.~A. {Muench}, and R.~J. {Elston}, \emph{\apj}
  \textbf{618}, 810--816 (2005).

\bibitem[{Muench} et~al.(2007)]{Muench07.1}
A.~A. {Muench}, C.~J. {Lada}, K.~L. {Luhman}, J.~{Muzerolle}, and E.~{Young},
  \emph{\aj} \textbf{134}, 411--444 (2007).

\bibitem[{Preibisch} et~al.(2005)]{Preibisch05.1}
T.~{Preibisch}, Y.-C. {Kim}, F.~{Favata}, E.~D. {Feigelson}, E.~{Flaccomio},
  K.~{Getman}, G.~{Micela}, S.~{Sciortino}, K.~{Stassun}, B.~{Stelzer}, and
  H.~{Zinnecker}, \emph{\apjs} \textbf{160}, 401--422 (2005).

\bibitem[{Grosso} et~al.(2007)]{Grosso07.1}
N.~{Grosso}, K.~R. {Briggs}, M.~{G{\"u}del}, S.~{Guieu}, E.~{Franciosini},
  F.~{Palla}, C.~{Dougados}, J.-L. {Monin}, F.~{M{\'e}nard}, J.~{Bouvier},
  M.~{Audard}, and A.~{Telleschi}, \emph{\aap} \textbf{468}, 391--403 (2007).

\bibitem[{Baraffe} et~al.(1998)]{Baraffe98.1}
I.~{Baraffe}, G.~{Chabrier}, F.~{Allard}, and P.~H. {Hauschildt}, \emph{\aap}
  \textbf{337}, 403--412 (1998).

\bibitem[{Chabrier} et~al.(2000)]{Chabrier00.2}
G.~{Chabrier}, I.~{Baraffe}, F.~{Allard}, and P.~{Hauschildt}, \emph{\apj}
  \textbf{542}, 464--472 (2000).

\bibitem[{Jardine} et~al.(2006)]{Jardine06.1}
M.~{Jardine}, A.~C. {Cameron}, J.-F. {Donati}, S.~G. {Gregory}, and K.~{Wood},
  \emph{\mnras} \textbf{367}, 917--927 (2006).

\bibitem[{Flaccomio} et~al.(2003)]{Flaccomio03.2}
E.~{Flaccomio}, G.~{Micela}, and S.~{Sciortino}, \emph{\aap} \textbf{402},
  277--292 (2003).

\end{thebibliography}

\IfFileExists{\jobname.bbl}{}
 {\typeout{}
  \typeout{******************************************}
  \typeout{** Please run "bibtex \jobname" to optain}
  \typeout{** the bibliography and then re-run LaTeX}
  \typeout{** twice to fix the references!}
  \typeout{******************************************}
  \typeout{}
 }

\end{document}